\documentclass[aps,showpacs, superscriptaddress,twocolumn,amsmath,amssymb,amsfonts]{revtex4-1}
\usepackage{amsmath}
\usepackage{graphicx}
\usepackage{bbm}
\usepackage{subfigure}
\usepackage{amssymb}
\usepackage{gensymb}
\usepackage{latexsym}

\usepackage{color}
 \usepackage[colorlinks,urlcolor=blue,linkcolor=blue,anchorcolor=blue,citecolor=blue]{hyperref}

\begin{document}

\title{Possible phases of the spin-$\frac{1}{2}$ XXZ model on a honeycomb lattice by boson-vortex duality}

\author{Han Ma}

\affiliation{Department of Physics, University of Colorado, Boulder, Colorado 80309, USA}
\affiliation{Center for Theory of Quantum Matter, University of Colorado, Boulder, Colorado 80309, USA}


\date{\today}

\begin{abstract}
Motivated by recent numerical work, we use the boson-vortex duality to study the possible phases of the frustrated spin-$\frac{1}{2}$ $J_1-J_2$ XXZ models on the honeycomb lattice.
By condensing the vortices,  we obtain various  gapped phases that  either break certain lattice symmetry or preserve all the symmetries.
The gapped phases breaking lattice symmetries occur when the vortex band structure has two minima. 
Condensing one of the two vortex flavors leads to an Ising ordered phase, while condensing both vortex flavors gives rise to a valence-bond-solid state. 
Both of those phases have been observed in the numerical studies of the $J_1-J_2$ XXZ honeycomb model. 
Furthermore, by tuning the band structure of vortex and condensing it at the $\Gamma$ point, we obtain a featureless paramagnet. But the precise nature of this featureless state is still unclear and needs future study.
\end{abstract}
\maketitle

\section{Introduction}

Frustration, the inability to simultaneously satisfy the competing interactions, can give rise to interesting physics.  
The study of  frustrated magnets can be traced back to the search for exotic phases of matter such as quantum spin liquids~\cite{anderson1973resonating}. 
There are two types of frustration, attributed to geometry and interaction. 
For example, geometric frustration arises from the kagome and pyrochlore lattice, on which exotic spin liquid states were discovered~\cite{savary2016quantum,zhou2017quantum}. 
Interaction frustration can also lead to exotic long-ranged entangled states, such as the Kitaev spin liquid on the honeycomb lattice~\cite{kitaev2006anyons}.

In this paper we focus on the spin-$\frac{1}{2}$ $J_1-J_2$ XXZ model on the honeycomb lattice, where the interaction frustration plays an important role.
The Hamiltonian is
\begin{equation}
\begin{aligned}
H &= J_1 \sum_{\langle i,j \rangle} (S^x_i S^x_j + S^y_i S^y_j +\alpha S^z_i S^z_j ) \\&+ J_2 \sum_{\langle \langle i,k \rangle \rangle} ( S^x_i S^x_k + S^y_i S^y_k +\alpha S^z_i S^z_k ) \label{eq:XXZ_spin}
\end{aligned}
\end{equation}
Although the lattice is bipartite, the competing nearest and next-nearest interactions give rise to interesting ground states, especially when $J_2$ is comparable with $J_1$. In the limit of $J_2/J_1 \rightarrow 0$, the ground state is an antiferromagnetic state with antiparallel spins on different sublattices.
At large $J_2/J_1$, two triangular sublattices are decoupled, and each of them has the $120^\circ$ states as ground states. For the intermediate $J_2/J_1$, due to strong frustration, previous studies suspect that the ground state is a spin liquid state ~\cite{varney2011kaleidoscope,varney2012quantum,carrasquilla2013nature,
Mulder2010spiral, mezzacapo2012ground,mosadeq2011plaquette,clark2011nature,gong2013phase,di2014spiral}. 
However, more and more evidence from recent numerical results shows that exotic spin liquids are unlikely to appear~\cite{zhu2013unexpected,bishop2014frustrated,li2014phase,ganesh2013deconfined,zhu2013weak,zhu2014quantum,pujari2015transitions,ferrari2017competition}. 
In the XY limit ($\alpha=0$), an unexpected $z$-direction Ising order is found when $0.22 \lesssim J_2/J_1 \lesssim 0.36$~\cite{zhu2013unexpected,bishop2014frustrated,li2014phase}. 
At the SU(2) point ($\alpha=1$), a valence bond solid (VBS) forms in a similar parameter region~\cite{ganesh2013deconfined,zhu2013weak,zhu2014quantum,li2014phase,pujari2015transitions,ferrari2017competition}. 
Additionally, the transitions between those gapped states and antiferromagnetic order seem to be direct transitions.
If those transitions are continuous,  they might be examples of quantum criticality beyond the Landau-Ginzburg paradigm~\cite{senthil2004deconfined,senthil2004quantum}. Although numerics gives us simple and clear results for both XY and SU(2) models on the honeycomb lattice, we still lack a theoretical understanding of those phase diagrams.

Besides those lattice-symmetry-breaking states, previous studies have shown that a short-range-entangled paramagnet, i.e., symmetric and nonfractionalized gapped groundstate, is also possible for the spin-$\frac{1}{2}$ system on the honeycomb lattice. The existence of such a featureless state is consistent with the Lieb-Schultz-Mattis theorem~\cite{lieb1961two} in  two dimensions~\cite{oshikawa2000commensurability,hastings2004lieb,hastings2005sufficient, parameswaran2013topological, watanabe2015filling, po2017lattice}. Although its wave function has been microscopically constructed~\cite{kimchi2013featureless,ware2015topological,jian2016existence,kim2016featureless}, the corresponding parent Hamiltonian is still unclear. Hence, it would be helpful to understand the physical mechanism for the featureless state in order to get its Hamiltonian.

In this work, we start with a generalized Bose-Hubbard model which could recover the above spin model in the particular limit. 
By carrying out the boson-vortex duality, we get an effective theory in terms of vortices coupled to a U(1) gauge field. 
The superfluid (i.e., the magnetic order in the spin language) corresponds to a state where the vortices are gapped. The vortices hopping on the triangular lattice have two low-energy modes at finite momenta ${\bf Q}_\pm=\pm(\frac{2}{3}\pi, \frac{2}{3}\pi)$. 
By condensing the vortex in various ways, the superfluid is disordered to the charge-density wave (CDW) order, VBS, and a featureless state. 
For example, if the vortices condense at one of the momenta ${\bf Q}_\pm$, we obtain the CDW (i.e., the Ising order in the spin language), which breaks the inversion symmetry on the bond of the honeycomb lattice. 
When vortices condense at both ${\bf Q}_\pm$, we obtain the valence-bond order, such as plaquette VBS (p-VBS) and columnar VBS (c-VBS). 
Finally, when vortices condense at the $\Gamma$ point $\left[{\bf Q} = (0,0)\right]$, a featureless state is obtained.

This paper is structured as follows. In Sec. \ref{sec:model}, we briefly review the boson-vortex duality on the honeycomb lattice. Then, we formulate the effective vortex theory on the triangular lattice in Sec. \ref{sec:dual_theory}. Accordingly, symmetry-breaking phases are obtained by condensing those vortices.
Further, in Sec. \ref{sec:featureless}, we  tune the band structure of vortices and show how to obtain a featureless state through vortex condensation. Finally, in Sec. \ref{sec:summary}, we summarize and discuss open questions.

\section{Model \label{sec:model}}

We consider a generalized Bose-Hubbard model on the honeycomb lattice with a half boson per site, whose Hamiltonian is
\begin{equation}
\begin{aligned}
H_{HB} &= H_0 +H_U \\
&=- t_1 \sum_{\langle i,j \rangle } (a_i^\dag a_j + H.c.)
+ \frac{U}{2} \sum_{i} n_i ( n_i -1) \\
&+ V_1 \sum_{\langle i,j \rangle} (n_i-\frac{1}{2})( n_j -\frac{1}{2})-  \tilde{\mu}  \sum_{i} n_i + \dots
\label{eq:boson_ham}
\end{aligned}
\end{equation}
where $a_i (a_i^\dag)$ is the boson annihilation (creation) operator on site $i$ of the honeycomb lattice, $n_i$ is the boson occupation number at site $i$, $\tilde{\mu}$ is the chemical potential, which is zero at half filling, $t_1>0$, and the dots ($\dots$) include the short-range farther neighbor hopping and interaction terms. Notice that $t_1 \rightarrow -t_1$ will not change the physics of this model because we can always carry out the transformation $a_i \rightarrow -a_i$ on one sublattice to cancel the sign change of $t_1$. 
This model is related to the spin model in Eq.~(\ref{eq:XXZ_spin}) in the following way. At infinite $U$, the bosons become hard-core bosons. At half filling, the boson system is equivalent to a spin system with spin-$1/2$ degrees of freedom on each site. Spins and bosons are related by the mapping $a_i \rightarrow S^-_i $, $a_i^\dag  \rightarrow S^+_i $, and $n_i \rightarrow S^z_i +\frac{1}{2}$. When there are only nearest- and next-nearest-neighbor hopping and interaction terms, we can map the boson model to a spin-$1/2$ $J_1 -J_2$ XXZ model in Eq.~(\ref{eq:XXZ_spin}) whose superexchange $J_i$ and $S^z_i S^z_j$ anisotropy $\alpha$ satisfy $J_{1,2}/2=t_{1,2}$ and $V_{1,2} = J_{1,2} \alpha$.

We proceed with the boson-vortex duality according to the standard procedure~\cite{dasgupta1981phase,nelson1988vortex,fisher1989correspondence,lee1989anyon}. First, we write the action of the Hamiltonian in Eq.~(\ref{eq:boson_ham}) after representing the bosons by rotor operators $\left[ \hat{\phi}_i, \hat{n}_j \right] =i \delta_{ij}$, where $n_i$ is the boson number and $\phi_i$ is its phase factor.  The action is also a function of imaginary time slice $\Delta \tau$. Then we proceed to the Villain representation. The nearest hopping term is written as
\begin{equation}
\exp \left[ t_1 \Delta \tau \cos (\Delta_\alpha \phi_i) \right] \rightarrow \sum_{\{L_{i\alpha}\}} \exp \biggl( -\frac{L_{i\alpha}^2}{2t_1 \Delta \tau} +i L_{i\alpha} \Delta_\alpha \phi_i \biggr) 
\end{equation}
where $L_{i\alpha}$ are integer variables living on the links of the direct honeycomb lattice, representing the current of bosons. $\Delta_\alpha$ is the discrete lattice derivative along the $\alpha$ direction: $\Delta_\alpha \phi_i = \phi_{i+\alpha} - \phi_i$. 

In the presence of the second-nearest-neighbor hopping, instead of adding those terms into the action, we renormalize the nearest neighbor hopping amplitude $t_1$. This treatment is allowed since in the end, all the parameters in the effective vortex theory are renormalized values~\cite{lannert2001quantum,balents2005putting}. Then, by integrating out the bosonic field $\phi$, we obtain the continuity equation for the bosonic three-current $J_{i\mu}= (n_i, L_{ix}, L_{iy})$, where $n_i$ is the boson density assigned to be along the time direction and $L_{i \mu}$ is the boson current along the spatial direction starting from site $i$. The continuity equation $ \Delta_\mu J_{i\mu}=0$ can be solved by defining a non-compact U(1) gauge field as
\begin{equation}
J_{j\mu} = \epsilon_{\mu\nu\lambda} \Delta_\nu A_{\mathcal{J}\lambda}. \label{eq:current_gauge}
\end{equation}
where $A_{\mathcal{J}\lambda}$ is a gauge field living on the dual triangular lattice link starting from dual lattice site $\mathcal{J}$ in the $\lambda$ direction. Below, we use lowercase letters like $i$, $j$, $k$ for the sites of the direct lattice, while the curly uppercase letters such as $\mathcal{J}$, $\mathcal{K}$, and so on are used for dual lattice sites. 

In terms of $J_{i\mu}$, we can also rewrite the other terms in Eq. (\ref{eq:boson_ham}) by changing $n_i$ to $J_{i\tau}$, for example, 
$ n_{i} (n_{i}-1)  = J_{i\tau}( J_{i\tau} -1)$.
We also absorb the nearest-neighbor interaction to the on-site interaction with a renormalized strength $\tilde{U}$. 
$\Delta \tau$ is chosen so that $e^2=t \Delta \tau= 1/\tilde{U}\Delta \tau$, where $t$ and $\tilde{U}$ are the renormalized hopping amplitude and on-site interaction of direct bosons.  

With all those ingredients, we are ready to derive the dual U(1) gauge theory. We will present it and study it carefully in the next section.

\begin{figure}[h]
\includegraphics[width=.3\textwidth]{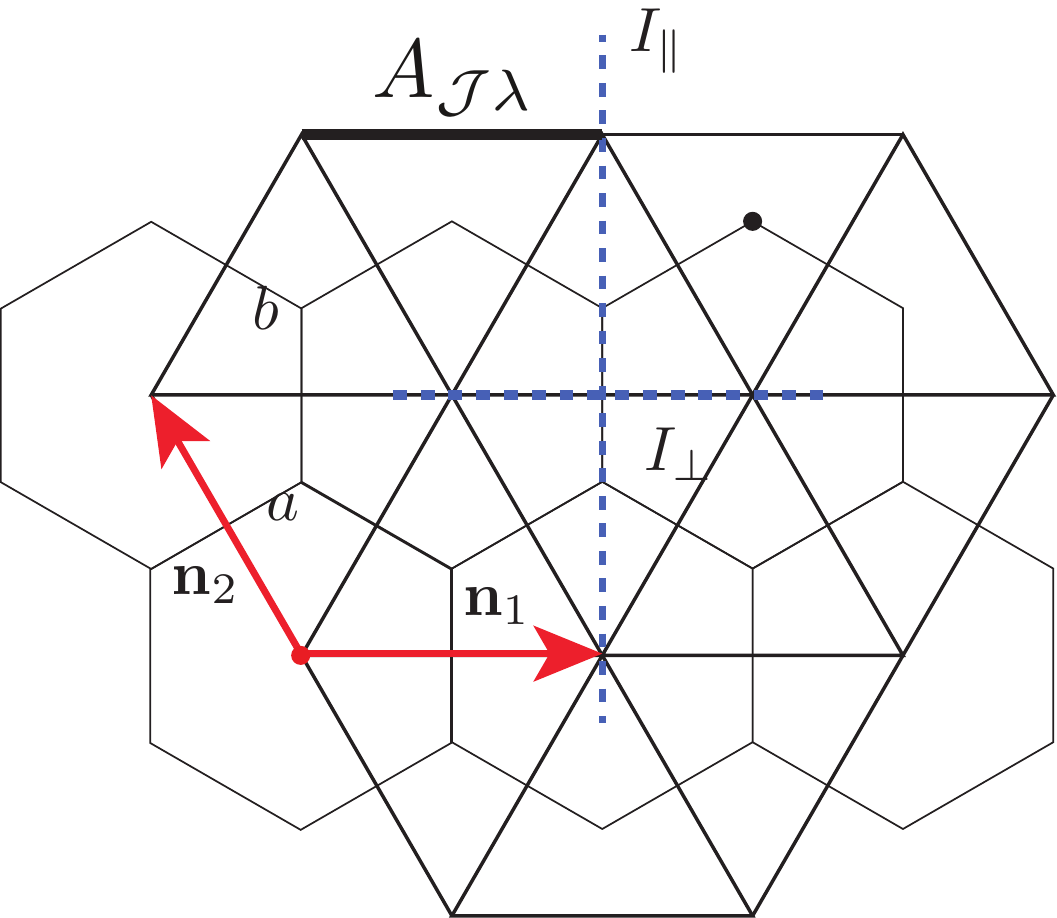}
\caption{Dual triangular lattice for the direct honeycomb lattice. $\vec{n}_1=(1,0)$ and $\vec{n}_2 =(-\frac{1}{2},\frac{\sqrt{3}}{2})$ are Bravais vectors spanning the dual lattice. $A_{\mathcal{J}\lambda}$ is the gauge field on the $\lambda$ bond, which is coupled to a vortex field on the $\mathcal{J}$ site. $I_\parallel$ and $I_\perp$ are inversion symmetries of the honeycomb lattice. \label{fig:triangle_lattice}}
\end{figure}

\section{Duality and  phases \label{sec:dual_theory}}

The dual vortex theory is defined on the triangular lattice in the Fig.~\ref{fig:triangle_lattice}. Its effective action describes a bosonic field, i.e., the vortex field, coupled to a noncompact U(1) gauge field living in the (2+1)-$d$ stacked triangular lattice, which is given by 
\begin{equation}
\begin{aligned}
\mathcal{Z} &=  \int \mathcal{D}A_{\mathcal{J}\lambda} \int \mathcal{D}\Psi_\mathcal{J} \exp \biggl\{  \sum_{\square \in P_{x\tau} \cup P_{y \tau}} \frac{-1}{2e^2}  (\epsilon_{\mu \nu \lambda} \Delta_\nu A_{\mathcal{J} \lambda} )^2 \\&-\sum_{\triangle \in P_{xy}}  \left[ \frac{1}{2e^2}  (\epsilon_{\tau \nu \lambda} \Delta_\nu A_{\mathcal{J} \lambda}-f )^2 \right] \\&-t_v \sum_{\mathcal{J}\lambda}(\Psi_\mathcal{J}^\dagger  e^{2\pi i A_{\mathcal{J}\lambda}} \Psi_{\mathcal{J}+\lambda} +H.c.) 
\\&-\sum_\mathcal{J} ( s|\Psi_\mathcal{J}|^2+\frac{u}{2} |\Psi_\mathcal{J}|^4) \biggr\}, \\ \label{eq:Z}
\end{aligned}
\end{equation}
where $t_v>0$ is the hopping amplitude of vortex $\Psi_\mathcal{J}$ coupling with gauge field $A_{\mathcal{J}\lambda}$ defined in Eq.~(\ref{eq:current_gauge}), $f= \frac{1}{2} $ is the boson filling factor, and $\tau$ is the direction of imaginary time, while other Greek letters like $\mu$, $\nu$ denote spatial directions $x$ or $y$. Notice the term in the first line sums over square plaquettes on the $x\tau$ and $y\tau$ planes denoted by $P_{x(y)\tau}$. The second line is a term summing over triangular plaquettes of the triangular lattice on the $xy$ plane. 

The second term of $\mathcal{Z}$ has a mean-field solution satisfying $\epsilon_{\tau\nu\lambda} \Delta_\nu A_{\mathcal{J}\lambda}=f$. This means a vortex sees a $\pi$ flux threading through each triangular plaquette due to the direct boson density. We choose a simple gauge with $A_{\mathcal{J}\lambda}=1/2$ on every link. After solving the band structure of vortices, we obtained the low-energy vortex field as a function of position ${\bf r}$,
\begin{equation}
\Psi ({\bf r})= \psi_1 e^{i {\bf Q}_+ \cdot {\bf r}} + \psi_2 e^{-i {\bf Q}_- \cdot {\bf r}},
\end{equation}
where $\bf Q_\pm$ is the minima of the vortex band structure.
Specifically, we introduce the basis vectors of the reciprocal lattice $\vec{b}_1=(1,\frac{1}{\sqrt{3}})$ and $\vec{b}_2= (0,\frac{2}{\sqrt{3}})$; then ${\bf Q}_\pm =\pm (\frac{2\pi}{3}\vec b_1 + \frac{2\pi}{3} \vec b_2)$.
Under lattice symmetries and the global U(1) symmetry, based on the vortex transformations under these symmetries in the Appendix, two low-energy modes of the vortex transform as follows: 
\begin{equation}
\begin{aligned}
T_1 \psi_{1(2)} &\rightarrow  \psi_{1(2)} e^{\pm i \frac{2\pi}{3}}, \\
T_2 \psi_{1(2)} &\rightarrow  \psi_{1(2)} e^{\pm i \frac{2\pi}{3}}, \\
R_{\pi/3}^{dual} \psi_{1(2)} &\rightarrow  \psi_{2(1)}, \\
R_{2\pi/3}^{direct} \psi_{1(2)} &\rightarrow \psi_{1(2)},  \\
I_\parallel \psi_{1(2)} & \rightarrow \psi_{1(2)}^\ast, \\
I_\perp \psi_{1(2)} & \rightarrow \psi_{2(1)}^\ast, \\
\mathcal{C} \psi_{1(2)} &\rightarrow \psi^\ast_{2(1)}, \\
U(1) \psi_{1(2)} & \rightarrow \psi_{1(2)} e^{i\alpha}. \label{eq:psg}
\end{aligned}
\end{equation}
Here, we list the transformation under translation symmetries $T_{1,2}$ along the $\vec{n}_{1,2}$ directions and rotational symmetries $R_{\pi/3}^{dual} (R_{2\pi/3}^{direct})$ around dual (direct) lattice sites. Also, inversion symmetries and charge conjugation are studied.

Finally, the action in terms of low-energy modes $\psi_{1,2}$ preserving all the symmetries is
\begin{equation}
\begin{aligned}
S &=\int d^3 x \biggl\{ \sum_{\mathfrak{s}=1,2} \biggl[ |(\partial_\mu - A_\mu) \psi_\mathfrak{s}|^2  +r  |\psi_\mathfrak{s} |^2  \biggr] + u  \sum_{\mathfrak{s}=1,2} (|\psi_\mathfrak{s} |^2)^2 \\&+ u_4 |\psi_1 |^2|\psi_2 |^2 +  v_c \left[  (\psi_1 \psi_2^\ast)^3 + (\psi_1^\ast \psi_2 )^3\right]\biggr\} \label{eq:eff_action}
\end{aligned}
\end{equation}
The possible phases in this theory are demonstrated below. 

\subsection{Superfluid}

When vortices are gapped, i.e., $r>0$, we preserve the dual U(1) gauge symmetry in the vortex vacuum. 
The photon mode of the U(1) gauge field can be identified as the Goldstone mode of the direct bosons.
This gives the superfluid state of direct bosons. By condensing vortices in different ways, we break the dual U(1) symmetry and restore the U(1) symmetry of direct bosons, resulting in gapped states with various orders.

\subsection{Gapped ordered state}

$r<0$ and $u>0$ lead to $\sum_{\mathfrak{s}=1,2} |\psi_\mathfrak{s}|^2 >0$, i.e., at least one of the two vortex flavors condenses. 

\subsubsection{Charge-density wave}

When $u_4>0$, single-vortex condensation is energetically favored. This means $\langle \psi_1 \rangle \neq 0,  \langle \psi_2 \rangle =0$ or, equivalently, $\langle \psi_1 \rangle = 0,  \langle \psi_2 \rangle \neq 0$. 
Suppose $\psi_1$ is condensed, i.e., $\psi_1 =  e^{i\theta_1}$. Without losing generality, we can set $\theta_1=0$. This leads to the vortex field
\begin{equation}
\Psi ({\bf r})= e^{i {\bf Q}_+ \cdot {\bf r} }.
\end{equation}
We can easily show that the translational symmetry is preserved, but $R_{\pi/3}^{dual}$, $I_\perp$, and $\mathcal{C}$ are all broken by this single-flavor condensation of $\psi_1$. But the combination $\mathcal{C} I_\perp$ is a symmetry of the resulting state.
Therefore, this phase is likely to be the CDW phase, which has a staggered boson density on the A and B sublattices.
This CDW phase is nothing but the unexpected Ising order that is discovered in the $J_1-J_2$ XY model~\cite{zhu2013unexpected}.

\begin{figure}[h]
\includegraphics[width=.45\textwidth]{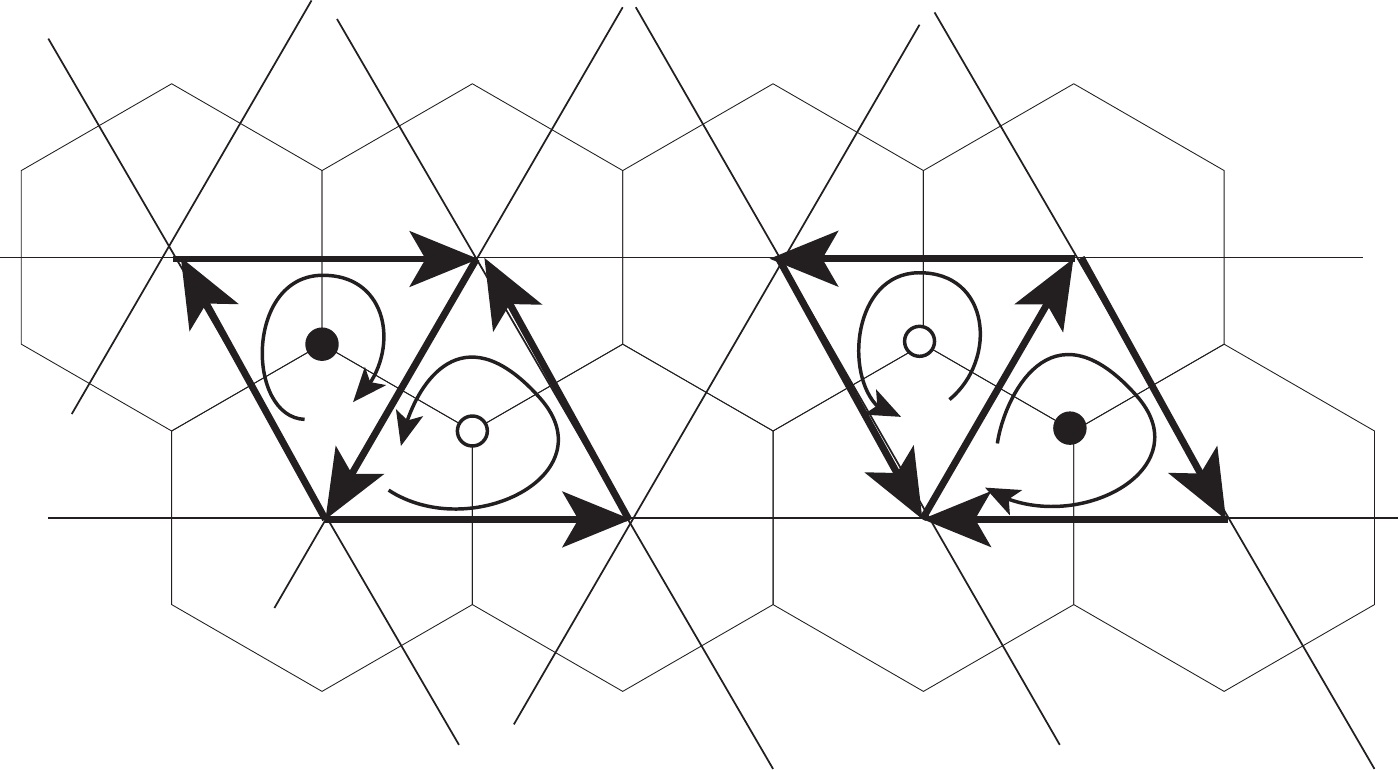}
\caption{Vortex current on the triangular dual lattice when the single vortex condenses. Two patterns of vortex current lead to two degenerate ground states of the CDW phase with lattice symmetry breaking of the direct boson. Specifically, the density on one sublattice (open circles) is higher than the other sublattice (solid circles).  \label{fig:vcurrent}}
\end{figure}

We can also show in a direct way that such single-flavor condensation leads to the CDW phase~\cite{lannert2001quantum,balents2005putting}.
We consider the vortex current, defined on the links of the dual triangular lattice as $J^v_{{\bf r},\mu} = i \Psi^\dag_{\bf r} D_{A,\mu} \Psi_{\bf r} $ where $D_{A,\mu} = \Delta_{\mu} - 2\pi i \vec{A}_\mu$ is the covariant lattice derivative on the links along the $\mu$ direction. The $2\pi$ in front of $A$ represents the unit of charge of the gauge field, which is also consistent with our convention in Eq. (\ref{eq:Z}).
As shown in Fig. \ref{fig:vcurrent}, the vortex currents $\vec{J}^v$ around the upward triangle and downward triangular plaquettes are opposite. The current also forms a vortex whose core lives at the center of each triangle.
 Since the vortex of the dual vortex corresponds to the direct boson, this pattern of the vortex current determines the boson density, and the corresponding state is the CDW state.

\subsubsection{Valence-bond-solid states}

When $u_4<0$, both species of vortices condense, i.e., $\langle \psi_1 \rangle =  \langle \psi_2 \rangle \neq 0$. Now the sign of $v_c$ determines the resulting ordered state.
Similarly, we assume $\psi_{1,2}  = e^{i \theta_{1,2}} $, and the effective Lagrangian is reduced to
\begin{equation}
\mathcal{L}_{eff}  = v_c \cos 3\theta,
\end{equation}
where $\theta = \theta_1-\theta_2$ is the difference in the phase factors of two species of vortices.
 
Restoring the boson U(1) symmetry, the resulting state is a gapped state. With the fixed $\theta$ determined by $\mathcal{L}_{eff}$, the ground state breaks translational symmetries $T_{1,2}$ and the $C_6$ rotational symmetry around the dual lattice sites down to $C_3$. Since both inversion symmetries and the charge conjugation are preserved, the corresponding state of matter should have bond order, where the singlet lives on certain bonds of the honeycomb lattice forming long-range order. There are two patterns of valence bonds satisfying this symmetry breaking. Different $v_c$ favor ground states with distinct bond patterns.

When $v_c <0$,  $\theta=\frac{2\pi}{3}n$ ($n \in \mathbb{Z}$) minimizes the action. The resulting vortex field is
\begin{equation}
\begin{aligned}
\Psi({\bf r}) &= e^{i{\bf Q}_+ \cdot {\bf r}+i\frac{n\pi}{3} } +e^{-i{\bf Q}_+ \cdot {\bf r}-i\frac{n\pi}{3}}  \\
&= 2 \cos \left[\frac{2\pi}{3} ( r_1 + r_2)+ \frac{n\pi}{3}\right] \\
\end{aligned}
\end{equation}
where $r_{1(2)}$ is the length of a component of ${\bf r}$ along the ${\bf n}_{1(2)}$ direction, which is shown in Fig.~\ref{fig:triangle_lattice}.

\begin{figure}[h]
\includegraphics[width=.48\textwidth]{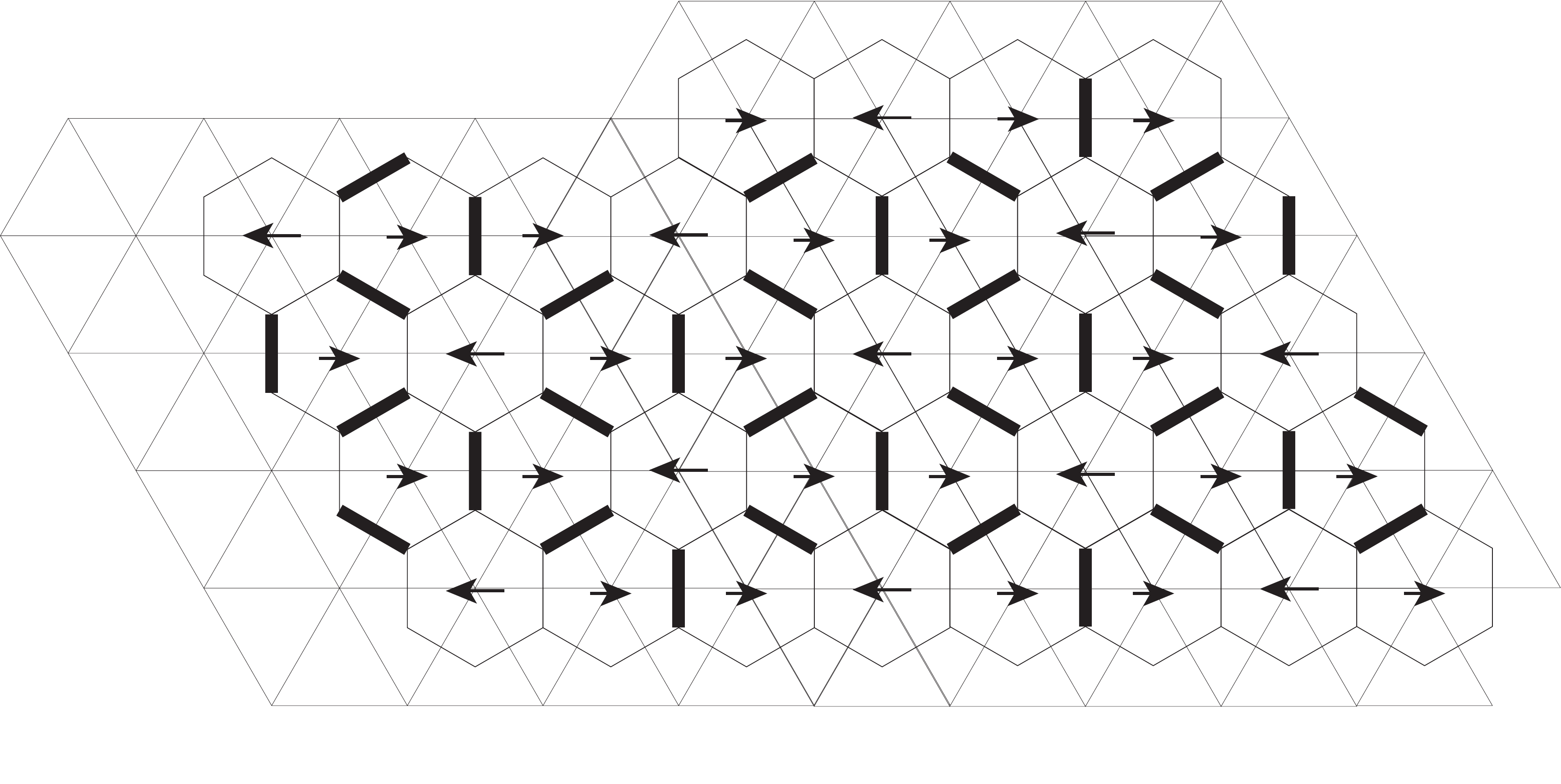}
\caption{The c-VBS of bosons, whose singlets are denoted by thick links. The corresponding vortex configuration is shown by arrows. \label{fig:vbs1}}
\end{figure}

We can calculate the vortex hopping amplitude $-\langle e^{i A_{\lambda}({\bf r})} \Psi^\ast ({\bf r}) \Psi({\bf r}+\lambda)\rangle$ to obtain the symmetry-breaking state of the direct bosons~\cite{lannert2001quantum,balents2005putting}.
Specifically, frustrated links of vortices reveal the locations of boson singlets. Notice $n =1,2,3$ leads to three different $\Psi$ configurations. They are the three degenerate states, one of which is shown in Fig. \ref{fig:vbs1}. 
We represent the vortex field $\Psi({\bf r})$ as arrows, whose length is proportional to $|\Psi|$ and whose direction represents the sign of $\Psi$. 
If two neighboring arrows are parallel (i.e. the two vortex fields have the same sign), the hopping between them will cost larger energy (since the expectation value of each hopping term is positive). 
Therefore, the vortex hopping is suppressed, and boson hopping across these frustrated links will be favored.
In other words, the bosons will form singlets on those links of direct honeycomb lattice.
Therefore, the c-VBS state (shown in Fig. \ref{fig:vbs1})  with threefold degeneracy is obtained. 

\begin{figure}[h]
\includegraphics[width=.48\textwidth]{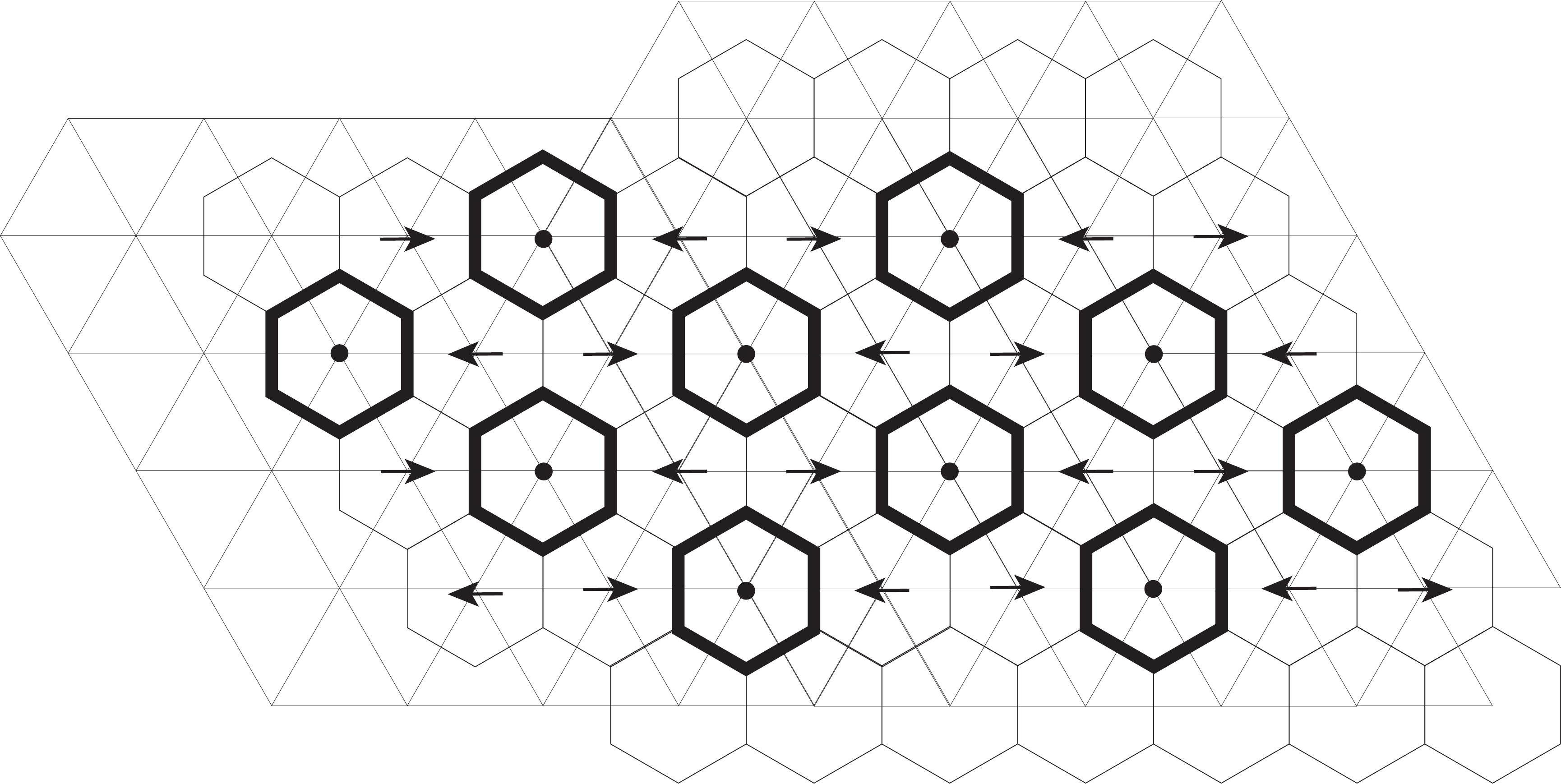}
\caption{p-VBS of bosons whose singlets are denoted by thick links. The corresponding vortex configuration is shown by arrows. Black dots denotes $\Psi = 0$ where the direction of the arrow is ambiguous. \label{fig:vbs2}}
\end{figure}

When $v_c >0$, then minimal action requires $\theta=\frac{\pi}{3} (2n+1)$ ($n \in \mathbb{Z}$). There are also three degenerate ground states for $n=0,1,2$. Now the vortex field is
\begin{equation}
\begin{aligned}
\Psi({\bf r}) &= e^{i{\bf Q}_+ \cdot {\bf r}+i\frac{\pi}{6}+i\frac{n\pi}{3}} +e^{-i{\bf Q}_+ \cdot {\bf r}-i \frac{\pi}{6}-i\frac{n\pi}{3} }  \\
&= 2\cos \left[ \frac{2\pi}{3} (r_1 + r_2) +\frac{\pi}{6}+\frac{n\pi}{3} \right]
\end{aligned}
\end{equation}
Also via the analysis of vortex frustrated links, we can get the ground state in terms of direct bosons, which has a plaquette order as shown in Fig.~\ref{fig:vbs2}. Notice that the c-VBS and p-VBS break the same lattice symmetry. Since both of them are product states, they belong to the same valence-bond phase. This means that when we tune $v_c$ from negative to positive, there is no phase transition at $v_c=0$. 

\begin{figure}[h]
\includegraphics[width=.4\textwidth]{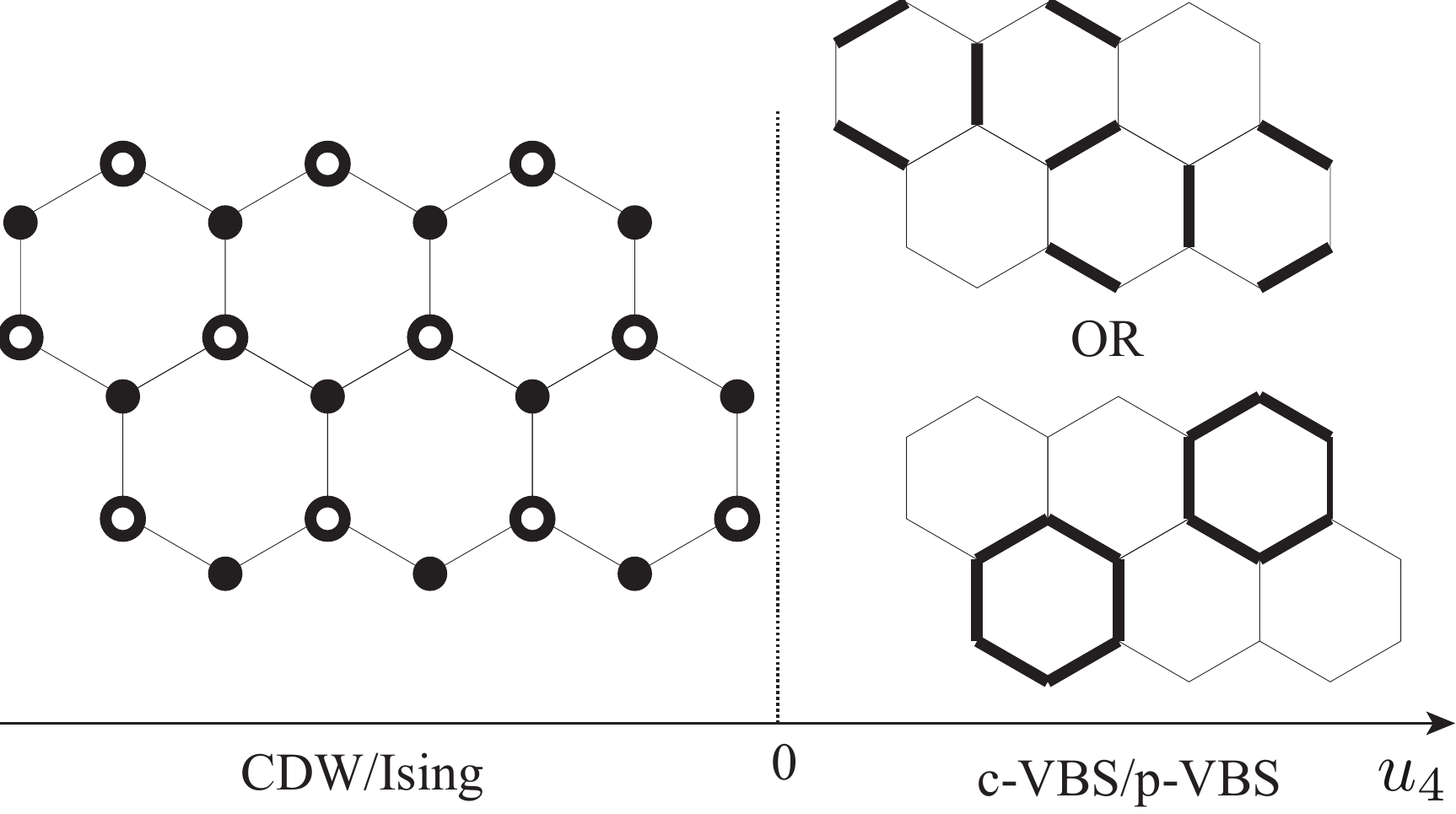}
\caption{Phase diagram obtained for $S$ in Eq. (\ref{eq:eff_action}) at $r<0$ and $u>0$. By tuning $u_4$, we get the CDW (Ising order in the spin language) for $u_4<0$ and the VBS state when $u_4>0$. The open circles and dots denote different densities of bosons occupied by different sublattices. \label{fig:phase_pheno} }
\end{figure}

In conclusion, we have found CDW and VBS states through vortex condensation. These results are summarized in Fig. \ref{fig:phase_pheno}.
Although we cannot find the exact relation between the microscopic boson model and the effective vortex theory, our phenomenological theory provides a physical mechanism for those states found in numerics.
This dual picture, on the other hand, allows us to construct the featureless state in terms of vortices, which will be studied below. 

\subsection{Featureless state \label{sec:featureless}}

Although the dual triangular lattice has a $\pi$ flux threading through each triangular plaquette, the unit cell of this dual lattice is not enlarged. Therefore, it is possible to have a single low-energy vortex mode carrying an integer quantum number of those symmetries, unlike Eq.~(\ref{eq:psg}). 
 Different from the square lattice~\cite{lannert2001quantum,balents2005putting}, the two minima of the vortex band structure on the triangular lattice are not protected by any symmetry.
 We thus can tune the band structure so that the only minimum locates at the $\Gamma$ point and the corresponding ground state is no longer degenerate. If the vortices condense at the $\Gamma$ point, we would have a state without lattice symmetry breaking. Since the gauge field is completely gapped out by the condensation, the resulting state is not a topological ordered state but a short-range entangled paramagnet. This is consistent with the Lieb-Shultz-Matthis theorem due to the integer boson per unit cell.

In order to get the featureless state, we tune the vortex band structure to have a minimum located at the $\Gamma$ point. 
With only the nearest-neighbor positive hopping, the energy minima locate at ${\bf Q}_\pm = \pm (\frac{2}{3}\pi \vec{b}_1+ \frac{2}{3}\pi \vec{b}_2)$ ($K$ points). As we discussed before, the condensation of vortices leads to degenerate ordered states. However, there is no reason to forbid farther-neighbor hopping. By adding those hopping terms, we can tune the vortex band structure and change the location of energy minima. 

\begin{figure}[h]
\includegraphics[width=.45\textwidth]{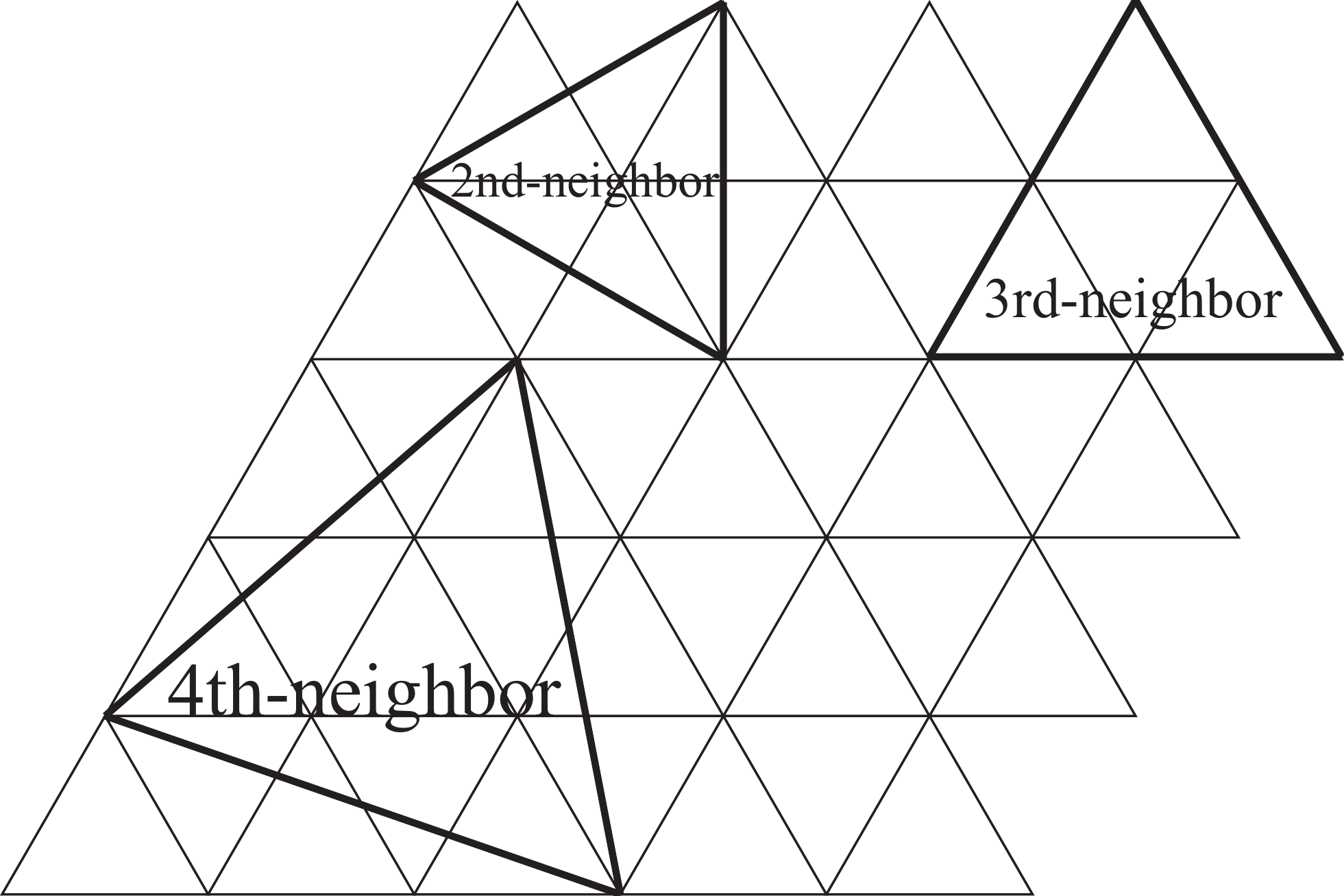}
\caption{Triangular plaquettes are formed by $n$-th neighbor hopping of vortices. \label{fig:triangle_NNN}}
\end{figure}

Notice each triangular plaquette must have a $\pi$ flux due to the $\frac{1}{2}$ boson per site. Then the triangular plaquette defined by the second-neighbor (NNN) hopping  (see Fig. \ref{fig:triangle_NNN}) also has a $\pi$ flux per plaquette because $\frac{3}{2}$ bosons are enclosed. If we consider only this NNN hopping of vortices, the band structure has three energy minima at $M$ points ${\bf Q} = \pi \vec{b}_1$, $\pi \vec{b}_2$, and $\pi (\vec{b}_1 + \vec{b}_2)$.

We find that if the third-neighbor hopping is the only vortex dynamics, the system has four degenerate ground states at the $\Gamma$
point and the $M$ points. This degeneracy is a fine tuning
effect and would be split by other hopping terms and interaction
terms. With appropriate interaction, it is possible to condense
the vortices at the point giving a symmetric short-range-entangled
state. The effective low energy vortex theory is the normal phi-4 theory. 
\begin{equation}
\mathcal{L} =| (\partial_\mu -A_\mu) \psi |^2 + r|\psi |^2 + u |\psi |^4
\end{equation}
Condensing vortices leads to a featureless Mott insulator. So far, it is not clear if it is a crystalline symmetry-protected topological state or just a trivial state which can be adiabatically connected to a product state. 

\section{summary and discussion \label{sec:summary}}

In this work, we studied the possible phases of the Bose-Hubbard model on the honeycomb lattice at half filling. 
Our study provides an approach to understand the phases found in the numerical study of the $J_1-J_2$ spin-$1/2$ XXZ models~\cite{zhu2013unexpected,zhu2013weak}.

By using the standard boson-vortex duality, we obtain a dual vortex theory on the dual triangular lattice. 
A state with gapped vortices corresponds to a superfluid phase. 
Then by condensing these vortices,  the superfluid is disordered to insulating phases.
The insulating phases can either break certain lattice symmetries or preserve all the symmetries, and it depends on the details of the vortex band structure.
Specifically, when the dynamics of the vortex is dominated by the nearest-neighbor hopping, the vortex band structure has two minima. 
Then the condensing vortex would necessarily break lattice symmetry, giving rise to CDW or VBS states. 
The CDW is obtained by condensing one of the two vortex flavors, and in the spin language, it corresponds to the Ising ordered state discovered numerically in the $J_1-J_2$ spin-$1/2$ XY honeycomb model~\cite{zhu2013unexpected}. 
The VBS state, including p-VBS and c-VBS patterns, is a condensate of two vortex flavors. 
The VBS state, particularly the p-VBS order, has been found in the numerical study of the spin-$1/2$ SU(2) $J_1-J_2$ honeycomb model~\cite{zhu2013weak}. 

Alternatively, if the vortex dynamics is dominated by the third-neighbor hopping, the vortex band structure is possible to have a single minimum at the $\Gamma$ point.
By condensing this single vortex flavor, we preserve all the lattice symmetries.
The existence of such a featureless state is consistent with the extended Lieb-Schultz-Mattis theorem in two dimensions. 
But the construction or realization of this state is nontrivial~\cite{kimchi2013featureless,ware2015topological,jian2016existence,kim2016featureless}. According to the present study, we need large third-neighbor hopping terms of vortices in the dual triangular lattice, which is unusual in realistic system.

Notice that there is another VBS state found in the numerical calculation, called a staggered VBS (s-VBS) state~\cite{zhu2013unexpected,zhu2013weak,ferrari2017competition}. Our theory cannot obtain this state directly from a superfluid. The reason is that the s-VBS state has a $Z_3$ vortex with a featureless core~\cite{xu2011quantum}. From the VBS side, by condensing this vortex, we cannot get a superfluid. Therefore, it is unlikely we will get a direct transition from the superfluid to the s-VBS state.

Due to the fact that the boson-vortex duality is a phenomenological theory which  cannot take into account all the microscopic details, we cannot predict the specific interaction which realizes those phases.
In particular the parent Hamiltonian of the featureless paramagnet is unclear.
Also we do not know whether the featureless paramagnet from our approach is the same phase as the one constructed in Refs. \cite{kimchi2013featureless,jian2016existence,kim2016featureless}, which is a crystalline symmetry-protected topological phase~\cite{ware2015topological}.

Moreover, the nature of the transitions between the superfluid (magnetic order) and those insulating phases is unknown. 
Numerically, it is unclear due to the finite-size effect. 
On the theoretical side, the phase transition between the superfluid and the lattice symmetry-breaking phase (CDW, VBS) is captured by the deconfined phase transition~\cite{senthil2004deconfined,senthil2004quantum}.
The transition from the superfluid to the CDW (Ising order) discovered in the  $J_1-J_2$ XY model is described by the easy-axis noncompact CP$^1$ (NCCP$^1$) theory~\cite{motrunich2004emergent}, which is likely to be first order,
while the transition from the superfluid to the p-VBS (or c-VBS) discovered in the  $J_1-J_2$  SU(2) model is described by the SU(2) NCCP$^1$ theory, which may be continuous.
Finally, the transition between the superfluid and the featureless paramagnet is naively described by an O(2) or O(3) Wilson-Fisher critical theory depending on the spin-rotational symmetry of the original spin model.

\acknowledgements

The author thanks Y. -C. He, M. Hermele, J. Y. Lee, S. Parameswaran, L. Radzihovsky and C. Wang for insightful discussions. The author was supported by M. Hermele's grant from the U.S. Department of Energy, Office of Science, Basic Energy Sciences (BES) under Award No.
DE-SC0014415.  This work was also supported in part by L. Radzihovsky's grant from the National Science Foundation under Grant
No. DMR-1001240 and by the Simons Investigator award
from the Simons Foundation. Part of this work was done at the Kavli Institute for Theoretical Physics, which is supported by the National Science Foundation under Grant No. NSF PHY11-25915 and a grant from the Gordon and Betty Moore Foundation.

\bibliography{./papernotes1}

\appendix*
\widetext

\section{Symmetry transformation of vortices on a triangular lattice \label{app:transf}}

Symmetry operations on the vortex operator $v(x,y)$ on the triangular lattice lead to
\begin{equation}
\begin{aligned}
T_1 v( n_1, n_2) &= v(n_1-1,n_2), \\
T_2 v( n_1, n_2) &= v(n_1,n_2-1), \\
R_{\pi/3}^{dual} v( n_1, n_2) &= v (n_1-n_2, n_1 ),  \\
R_{2\pi/3}^{direct} v(n_1 - \frac{1}{3}, n_2-\frac{2}{3}) &= v(-n_2 +\frac{2}{3} , n_1-n_2 +\frac{1}{3}), \\
I_\parallel v(n_1, n_2) &= v^\ast (n_2 -n_1, n_2), \\
I_\perp v(n_1,n_2) &= v^\ast (n_1-n_2, -n_2).
\end{aligned}
\end{equation}
Fourier transformation gives
\begin{equation}
\begin{aligned}
T_1 v(k_1, k_2) &= v(k_1,k_2) e^{i k_1}, \\
T_2 v(k_1, k_2) &= v(k_1,k_2) e^{i k_2}, \\
R_{\pi/3}^{dual} v(k_1, k_2) &=  v (-k_2, k_1 +k_2 ), \\
R_{2\pi/3}^{direct} v(k_1,k_2)  &= v(-k_1-k_2,k_1),  \\
I_\parallel v(k_1, k_2) &= v^\ast(k_1, -k_1-k_2), \\
I_\perp v(k_1, k_2) &= v^\ast(-k_1, k_1+k_2).
\end{aligned}
\end{equation}
Thus, for a particular momentum ${\bf Q}$, e.g. ${\bf Q}_\pm$ in this context, we obtain the corresponding transformations for low-energy vortex modes as listed in Eq. (\ref{eq:psg}).

\end{document}